&latex209 
\documentclass[11pt]{article}
\usepackage{moriond-photos,epsfig}
\usepackage{floatflt}
\usepackage{xspace}
\usepackage{amsmath}
\usepackage{amssymb}
\def\Journal#1#2#3#4{{#1} {\bf #2}, #3 (#4)}

\def\NPB{{\em Nucl. Phys.} B}
\def\PLB{{\em Phys. Lett.}  B}

\def\PRD{{\em Phys. Rev.} D}

\def\be{\begin{equation}}
\def\ee{\end{equation}}
\def\bea{\begin{eqnarray}}
\def\eea{\end{eqnarray}}

\begin{document}
\vspace*{4cm}
\title{CHARM PRODUCTION IN {\large $\mathbf{e p} $} INTERACTIONS AT HERA \\ and \\ EVIDENCE FOR A NARROW ANTI-CHARMED BARYON STATE AT H1}

\author{ K. LIPKA \\(on behalf of H1 and ZEUS Collaborations)}

\address{DESY Zeuthen, Platanenalle 6,\\
15738 Zeuthen, Germany}
\maketitle\abstracts{
Recent results on open charm production at HERA are presented. Charm quarks 
are identified via the reconstruction of D-mesons. The charm contribution to the proton structure 
function is shown. Evidence for an exotic anti-charmed baryon state observed by H1 is presented. 
The data show a narrow resonance in the $D^*p$ invariant mass combination at 
3099$\pm$3$_{stat}$$\pm$5$_{syst}$  MeV. The resonance is interpreted as an anti-charmed baryon with 
minimal constituent quark content $uudd\bar{c}$ together with its charge conjugate. Such a signal is not observed in a similar preliminary ZEUS analysis.}
\vspace*{-0.5cm}
\section{Open charm production in $ep$ collisions at HERA}
\begin{figure}[h]
  \begin{minipage}[l]{0.48\textwidth}\vspace{-30pt}
    \epsfig{figure=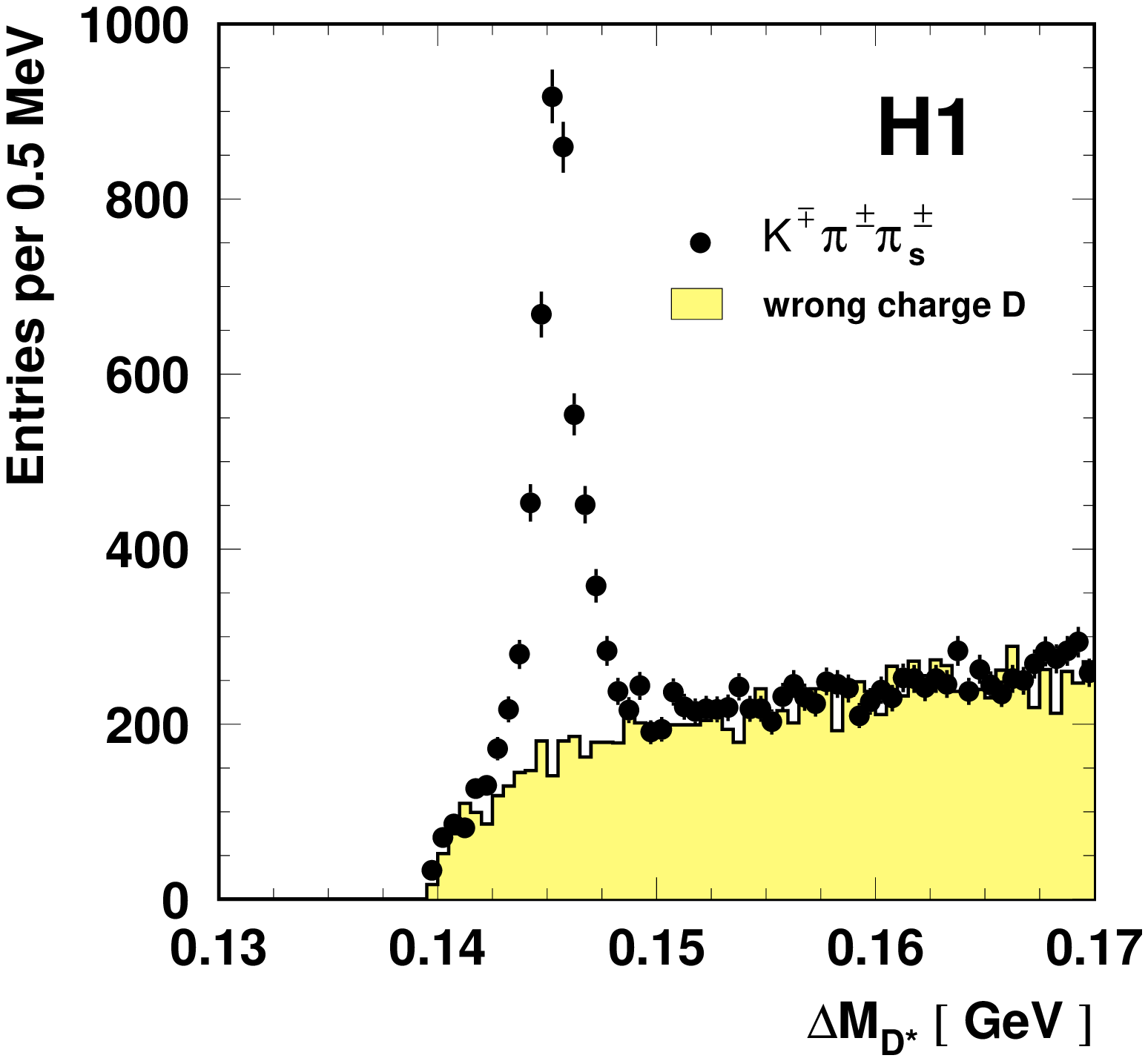,width=\textwidth}\vspace{-3pt} 
    \caption{$\Delta M_{D^*}$ distribution for $K^\mp \pi^\pm \pi_s^\pm$ combinations. Non-charm background, represented by the ``wrong charge D'' distribution obtained from $K^\pm \pi^\pm \pi_s^\mp$ is also shown.  
     \label{fig1}}
  \end{minipage}
  \hfill
  \begin{minipage}[r]{0.48\textwidth}\vspace{10pt}
    \epsfig{figure=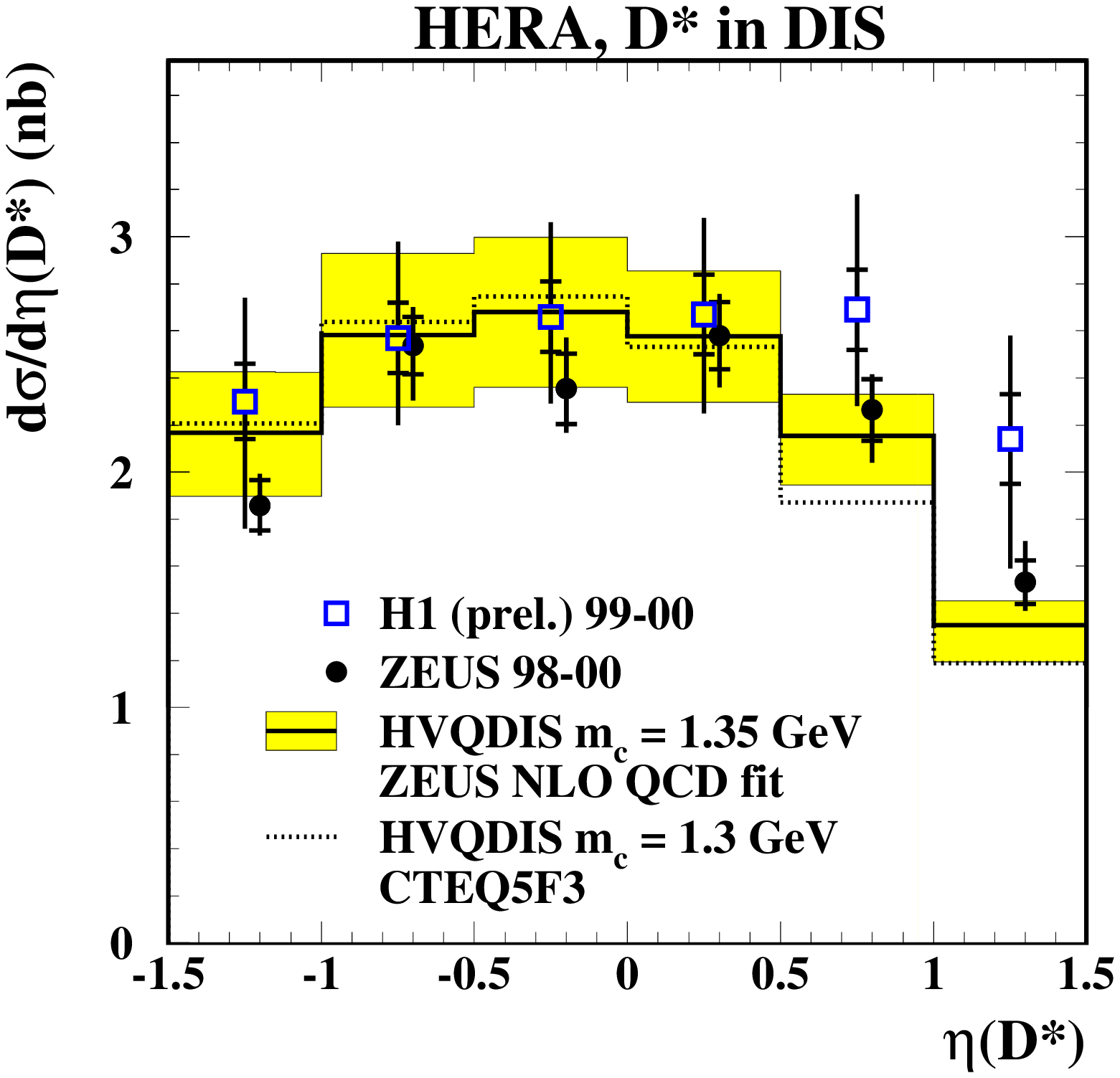,width=0.8\textwidth}
    \bigskip
    \caption{Differential $D^*$ cross section in DIS in bins 
of pseudorapidity $\eta_{D^*}$ compared to NLO QCD calculations (solid and dashed lines). The
band represents the theoretical uncertainty described in the text.
      \label{fig2}}
  \end{minipage}
\end{figure}

At HERA  27.5 GeV electrons collide with protons of 920 GeV yielding a center of mass energy of 318 GeV. 
In $ep$ interactions charm and anti-charm quarks are produced predominantly in boson-gluon fusion. 
The kinematics of $ep$ scattering are described by the virtuality of the exchanged photon 
$Q^2$, the Bjorken scaling variables $x$ and $y$ and the invariant mass of the photon-proton system $W$. 
Depending on the value of $Q^2$ two different kinematic regimes are exploited: the deep inelastic scattering (DIS) 
regime is characterised by $Q^2>$1 GeV$^2$, while in the photoproduction regime ($Q^2<$1 GeV$^2$) 
the electron escapes the main detector since it is scattered under very small angles.
Open charm production is tagged via $D^*$ production detected through its decay~\footnote{The notation $\pi_s$ is used to distinguish the low momentum pion released in $D^*$ decay from that from $D^0$ decay.} channel $D^{*\pm} \to D^0 + \pi_s^{\pm} \to K^\mp \pi^\pm + \pi^\pm_s$.
For the $D^*$ selection~\cite{h1f2c} the mass difference technique is used, based on variable
\vspace*{-0.3cm}
\begin{equation}
\Delta M_{D^*} = m(K \pi \pi_{s}) - m(K \pi)
\vspace*{-0.cm}
\end{equation} 
where $m(K \pi \pi_{s})$ and $m(K \pi)$ are the invariant masses of the corresponding combinations.
In Fig. \ref{fig1} an example of the $\Delta M_{D^*}$ distribution is shown for the DIS regime. A prominent 
signal on a modest background is seen around the expected $M(D^*)-M(D^0)$ mass difference. The distribution 
is compared with ``wrong charge D'' background where the $D^0$  is replaced by fake ``$D$-mesons'' composed 
of like-charge $K \pi$ combinations.
In Fig.~\ref{fig2} the measured~\cite{h1zeusxsec,h1zeusf2c} differential cross section of $D^*$ production 
is shown as a function of the pseudo-rapidity $\eta_{D*}=-ln(tan (\theta_{D*}/2))$ together with the NLO QCD
calculation~\cite{hvqdis}. Good agreement is observed between data and theoretical expectation. 
\begin{floatingfigure}{0.54\textwidth}
  \mbox{}\hspace{0.01\textwidth}\epsfig{file=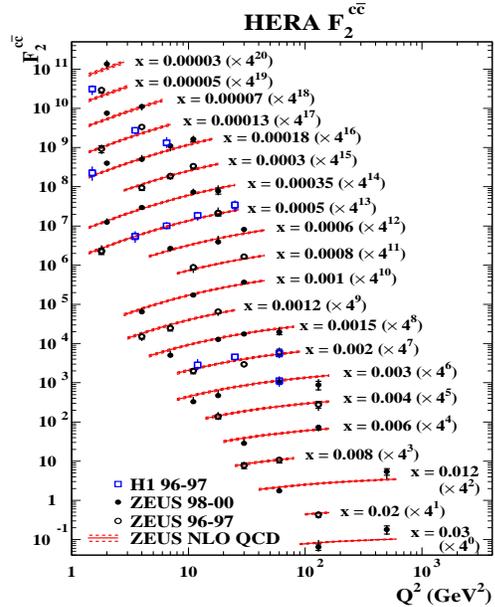,
    width=0.4\textwidth, height=0.5\textwidth} 
  \caption{Charm contribution $ F_2^{c \bar c}$ to the proton structure function ($F_2$) 
compared to the NLO QCD fits to inclusive data. \label{fig3}}
\end{floatingfigure}
\hspace*{-0.7cm} The uncertainties of the calculation are due to variation of the mass of the charm quark, 
factorisation and renormalisation scales and fragmentation parameters. Calculations using two sets of parton 
densities in the proton are shown.

The charm contribution to the proton structure function $ F_2^{c \bar c}$ was extracted~\cite{h1zeusf2c,h1f2c} from 
the DIS cross section of $D^*$ production. $ F_2^{c \bar c}$ is shown in Fig. \ref{fig3} in bins of $Q^2$ and 
$x$ and compared to the predictions from NLO QCD fits to inclusive measurements. 
The effect of scaling violations is 
clearly visible in the rise of $ F_2^{c \bar c}$ with $Q^2$ which becomes steeper with decreasing $x$.
These data will be used in the future to constrain the gluon density in the proton. The ratio of $F_2^{c \bar c}$ 
and the proton structure function $F_2$ rises to about 0.35 at large $Q^2$ and low $x$.
\newpage
\section{Evidence for a narrow anti-charmed baryon state at H1}
\begin{figure}[h]
\vspace*{-0.5cm}
  \begin{minipage}[l]{0.48\textwidth}\vspace{-35pt}
\hspace*{-0.4cm}
    \epsfig{figure=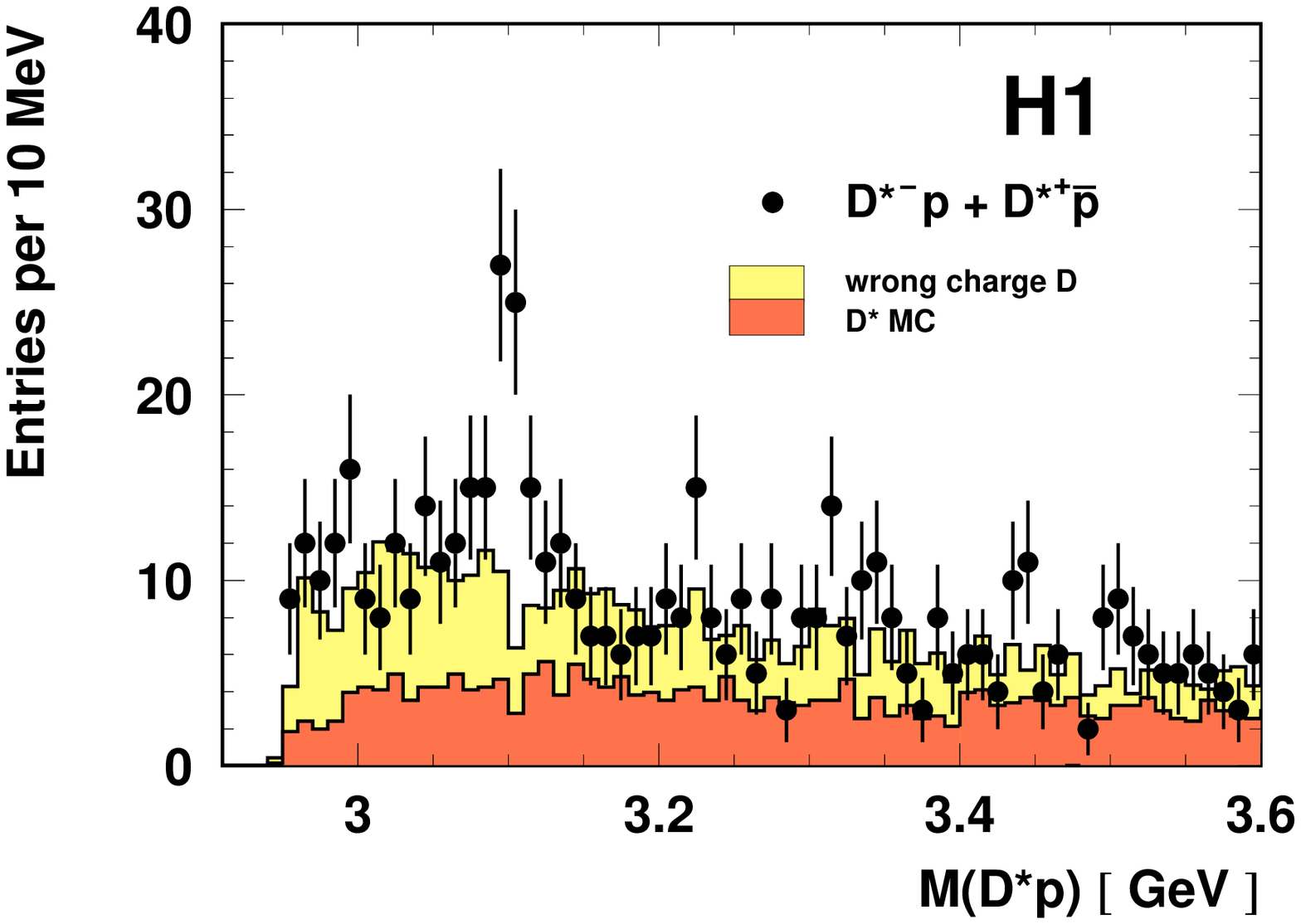,width=1.1\textwidth}\vspace{-3pt} 
    \caption{M($D^*p$) distribution for opposite-charge $D^*p$ combinations. The non-charm ``wrong charge D'' background distribution, and $D^*$ MC simulation are also shown.  
     \label{fig4}}
  \end{minipage}
  \hfill
  \begin{minipage}[r]{0.48\textwidth}\vspace{-45pt}
\hspace*{-0.4cm}
    \epsfig{figure=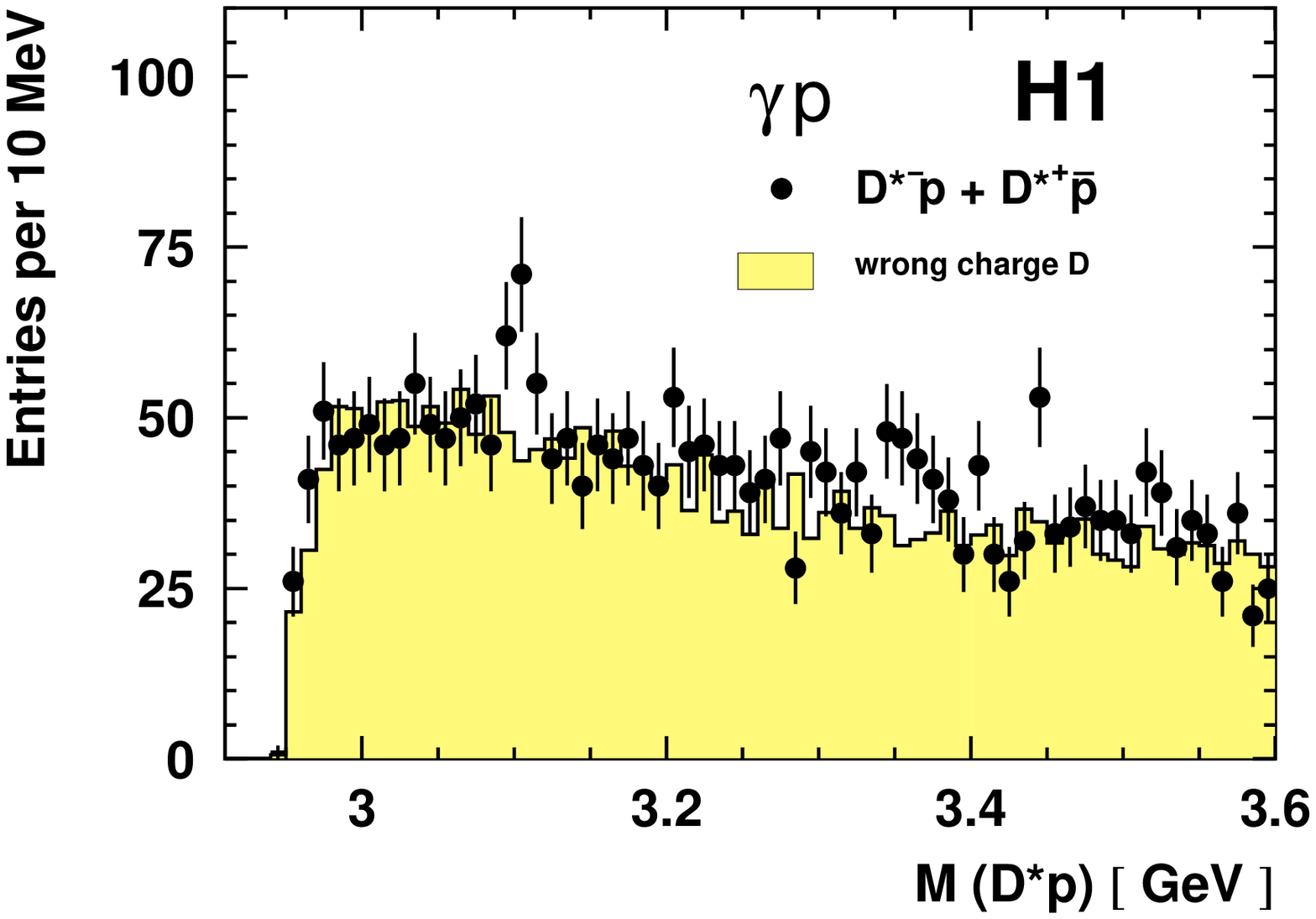,width=1.12\textwidth}
	\vspace*{-0.8cm}	
    \caption{M($D^*p$) distribution from the photoproduction analysis compared with a background model derived from ``wrong-charge D'' combinations. 	
      \label{fig5}}
  \end{minipage}
\end{figure}

A narrow resonance was observed in the mass spectrum of $D^{*-}p$ and $D^{*+}\bar{p}$ 
combinations, as recently published~\cite{h1cpq} by H1. 
For the analysis at H1 the DIS data of the running period 1996-2000 has been analysed, corresponding 
to a luminosity of 75 pb$^{-1}$. $D^*$ candidates were selected similarly to the $F_2^{c \bar c}$ analysis with 
additional requirements 
to further reduce the non-charm induced background. Only those candidates having a  $\Delta M_{D^*}$
value in a window $\pm$2.5 MeV around the nominal $M(D^*)-M(D^0)$ mass difference are combined with 
proton candidates.
The latter are selected with dE/dx requirements to suppress the pion and other background. 
The average dE/dx resolution at H1 is about 8\%.
The mass of the $D^*p$ state is calculated as 
\begin{equation}
M(D^*p) = m(K \pi \pi_{s} p) - m(K \pi \pi_{s}) + M_{PDG}(D^*)
\end{equation}
where $m(K \pi \pi_{s} p)$ and $m(K \pi \pi_{s})$ are the invariant masses of the corresponding particle 
combinations  to which the $D^*$ mass~\cite{pdg} $M_{PDG}(D^*)$=2010.0 MeV is added. A clear narrow peak is observed in the mass distribution as 
shown in Fig.\ref{fig4}. 
The data are compared with the $D^*$ Monte-Carlo (MC) simulation and ``wrong charge D'' background model. 
No enhancement is seen either in MC or in the non-charm background from data, while the shape 
of the background is very well described. The signal is seen separately in $D^{*-}p$ and $D^{*+}\bar{p}$ 
combinations with compatible significance, mass position and width. No significant enhancement is observed 
in like-charge $D^*p$ combinations. 

Possible kinematic reflections that could fake the signal have been ruled 
out by studying invariant mass distributions and correlations involving the $K, \pi, \pi_s$
and proton candidates under various particle mass hypotheses. 
All events in the M($D^*p$) distribution have been scanned visually. No anomalies are observed in the 
reconstruction of the candidate tracks.

\begin{figure}[h]
  \begin{minipage}[l]{0.48\textwidth}\vspace{-50pt}\hspace*{-0.5cm}
    \epsfig{figure=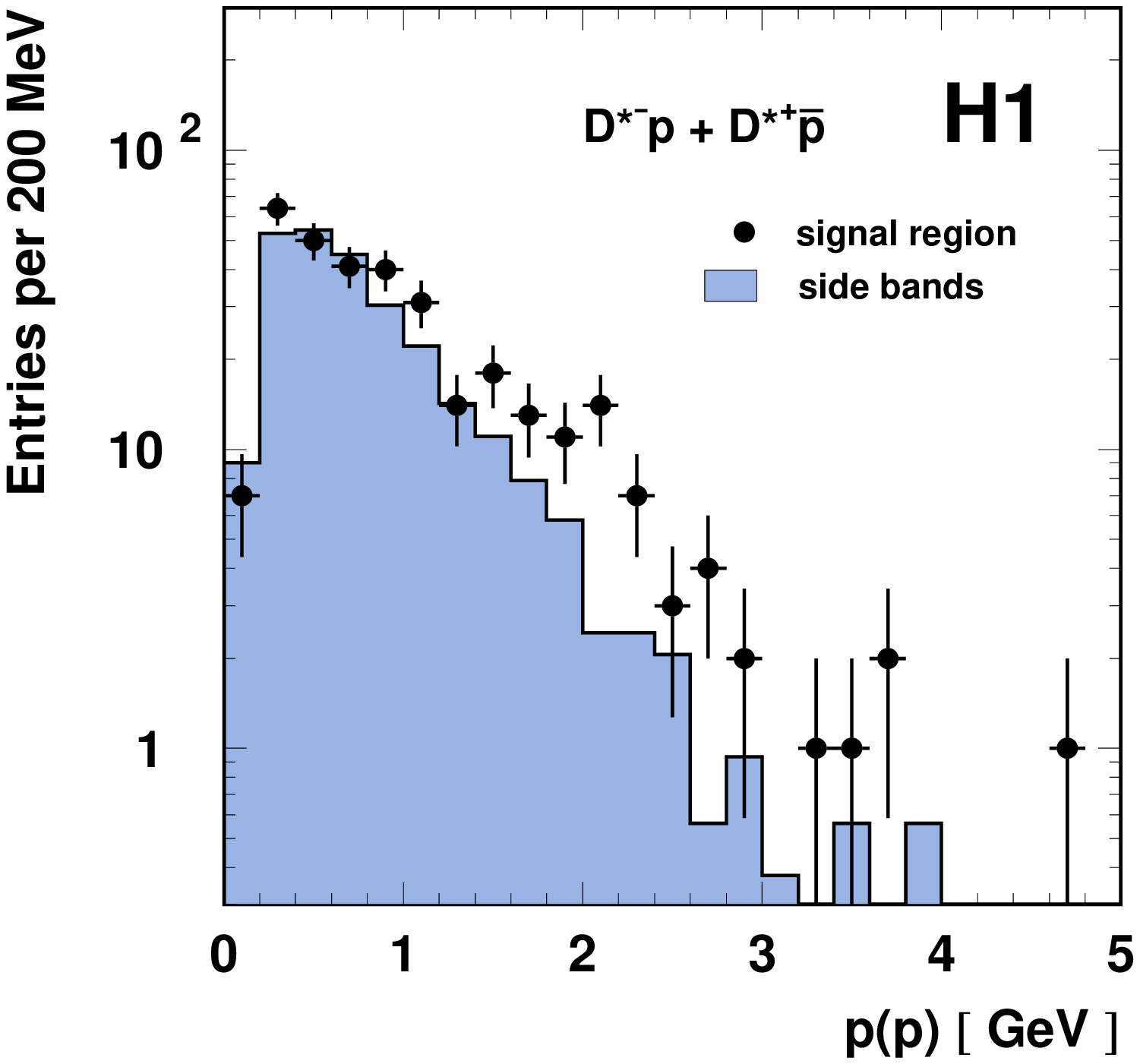,width=1.2\textwidth,height=1.1\textwidth}
	\vspace{-25pt}	
    \caption{Momentum distribution for charged particles yielding M($D^*p$) values falling in the signal 
and side band regions.  
     \label{fig6}}
  \end{minipage}
  \hfill
  \begin{minipage}[r]{0.48\textwidth}\vspace{-65pt} \hspace*{-1cm}
    \epsfig{figure=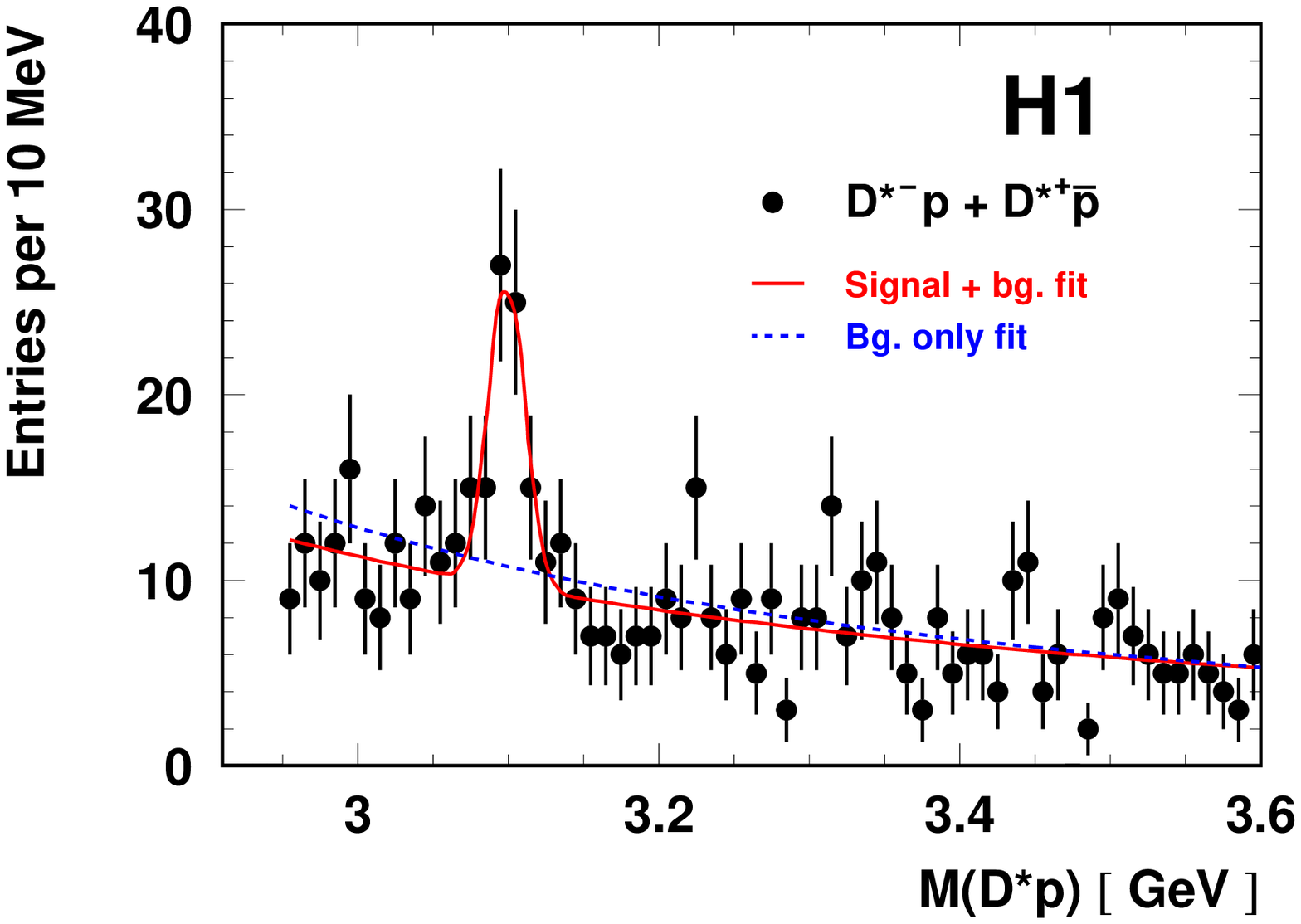,width=1.2\textwidth, height=1.33\textwidth}
	\vspace*{-30pt}
    \caption{$M (D^*p)$ distribution for opposite-charge $D^*p$ combinations in DIS compared 
with the results of the fits with both signal plus background components (solid line) and background only
component (dashed line). 
      \label{fig7}}
  \end{minipage}
\end{figure}

Several studies were performed to test the $D^*$ and proton content of the signal. 
It was shown that the $D^*p$ signal region is enriched with $D^*$ in comparison to the side bands. 
The signal is visible for low momentum proton candidates where protons can be unambiguously identified.
The signal is also observed in the independent photoproduction sample as shown in Fig. \ref{fig5}. 
The momentum distribution of the proton candidates without any dE/dx requirement shown in 
Fig. \ref{fig6} reveals a significantly harder spectrum in the $D^*p$ signal region compared 
to the sidebands. This supports the expected change in the $D^*p$ kinematics. 

The fits to the M($D^*p$) distribution in DIS are shown in Fig. \ref{fig7}. The measured width is 
consistent with the experimental resolution, therefore a Gaussian distribution is 
used for the signal shape yielding r.m.s. of 12$\pm$3$_{stat}$. The background is parameterised with 
a power law. 
The mass of the resonance is determined to be 3099$\pm$3$_{stat}$$\pm$5$_{syst}$ MeV.
The probability that the background distribution fluctuates to produce the signal is calculated 
considering the background-only hypothesis (dashed line in Fig.\ref{fig7}) to be less than 
4$\times$10$^{-8}$ which corresponds to 5.4 $\sigma$ in terms of Gaussian standard deviations.
A state decaying strongly to $D^{*-}p$ must have baryon number +1 and charm -1 and thus has a minimal 
constituent quark composition of $uudd\bar{c}$. Therefore the observed resonance is a candidate 
for the charmed pentaquark.

A similar preliminary search has been performed at the ZEUS experiment in
both photoproduction and DIS. Similar criteria were applied to select the
$D^*$ mesons and $dE/dx$ measurements were used to select proton
candidates. Data from the years 1995-2000 were analysed, corresponding
to an integrated luminosity of $126 \ {\rm pb^{-1}}$. No signal was
observed in either DIS or photoproduction.


\section*{References}
\begin{thebibliography}{99}
\bibitem{h1f2c} H1 Coll., \Journal{\PLB}{528}{199}{2002}.
\bibitem{h1zeusxsec} H1 Coll.,  {\it Abstract 098, International Eur. Conf. on HEP}, July 17-23, 2003;
\bibitem{h1zeusf2c} ZEUS Coll., DESY-03-115, accepted by {\it Phys. Rev. D}. 
\bibitem{hvqdis} B.W. Harris, J. Smith, \Journal{\NPB}{452}{109}{1995}
\bibitem{h1cpq} H1 Coll., [hep-ex/0403017], accepted by {\it Phys. Lett. B}. 
\bibitem{pdg} Particle Data Group, H. Hagiwara et al., \Journal{\PRD}{66}{010001}{2002}
\end{thebibliography}
\end{document}